\documentclass[prd,
eqsecnum,floatfix,letterpaper,showpacs,superscriptaddress,groupedaddress,nofootinbib]{revtex4-2}

\usepackage{changes}

\def \vss{\vspace{14pt}}

\usepackage{amsmath}
\usepackage{latexsym}
\usepackage{amssymb}
\usepackage{amsfonts}
\usepackage{mathtools}
\usepackage{bm}
\usepackage{amsthm}
\usepackage{graphicx}
\usepackage{color}
\usepackage{changes}
\usepackage[normalem]{ulem}
\usepackage{natbib}
\usepackage{appendix}
\usepackage{subcaption}
\def\no{\nonumber}

\def \omg {\omega}
\def \om0{\omega_0}
\def \Dom{\Delta \omega}
\def \F {\mathcal{F}}

\def \th{\tilde h}
\def \qtr {\frac{1}{4}}
\def \Msun {M_\odot}

\def \be{\begin{equation}}
\def \bea{\begin{eqnarray}}
\def \eea{\end{eqnarray}}
\def \ee{\end{equation}}

\begin{document}
\title{Detection of Gravitational Wave modes in third generation detectors}

\author{Massimo Tinto}
\email{massimo.tinto@gmail.com}
\affiliation{Divis\~{a}o de Astrof\'{i}sica, Instituto
  Nacional de Pesquisas Espaciais, S. J. Campos, SP 12227-010, Brazil}
 \author{Sanjeev Dhurandhar}
\email{sanjeev@iucaa.in}
\affiliation{Inter University Centre for Astronomy \& Astrophysicss, Ganeshkhind, Pune 411007, India}
\author{Harshit Raj}
\email{harshitrajharsh@gmail.com}
\affiliation{Indian Institute of Science Education and Research, Pashan, Pune 411008, India}
\date{\today}

\begin{abstract}
 We investigate the detectability of Gravitational Wave (GW) modes (emitted by black-holes and neutron stars) by third generation, ground-based gravitational wave detectors planned to be operational in the next decade. Our analysis focuses on the Cosmic Explorer and Einstein Telescope projects, which are expected to have arm lengths of tens of kilometers and to experience the amplification of a gravitational wave signal at their Full-Spectral Range (FSR) frequencies. We find that both projects will also observe with good Signal-to-Noise ratio (SNR) the elusive {\it w-modes}, which are expected to be emitted at these frequencies by spinning neutron stars.
\end{abstract}

\maketitle

\section{Introduction}
\label{SecI}

Gravitational wave (GW) astronomy has now become an established field of research with the successful development and continuous operation of laser interferometer detectors. The first generation interferometers, LIGO, Virgo, and GEO600, became operational during the first decade of the century when they proved their design sensitivities and operational capabilities. Second-generation detectors currently online, such as Advanced LIGO (aLIGO) \cite{LIGO}, Advanced Virgo (aVirgo) \cite{VIRGO} and KAGRA \cite{KAGRA}, have made groundbreaking discoveries since becoming operational, observing about 90 events during the so called O1 to O3 science runs and generating about 200 alerts during the last science campaign O4. In the first half of the O4 run, that is, O4a, 128 candidate events were found with $p_{\rm astro} > 0.5$ \cite{GWTC-4.0} ($p_{\rm astro} > 0.5$ means that the likelihood of finding a signal in the data is higher than the likelihood of data containing only noise). Current generation interferometers have also allowed the beginning of multi-messenger astronomy \cite{mlti-msngr2025}, the entirely new field of astronomy covering complementary observations in the electromagnetic and GW spectra, which promises to further enhance our understanding of the Universe. As future space-based interferometers, such as LISA, TaiJi and TianQin \cite{LISA2017,Taiji, TianQin} will allow us to make GW observations in the mHz frequency region, the possibility of also developing  ground-based interferometers of arm-lengths up to ten times longer than currently operational detectors will allow us to probe post-merger dynamics, detect continuous signals from isolated pulsars, and a stochastic GW background. At design sensitivity the current detectors will be able to observe neutron star mergers up to 200-250 Mpc on average at a SNR of 8. They  have so far detected predominantly black hole mergers and very few events containing neutron stars. This is a selection effect because the GW signals from black holes are much stronger than those of neutron stars because the black hole masses are much larger. Therefore, black hole signals can be observed from larger distances which results in a much larger event rate. The Einstein Telescope (ET) \cite{ET,ET_1,ET_2} and the Cosmic Explorer (CE) \cite{CE_1,CE_2} projects are expected to reach distances several times greater than those reached by the current advanced detectors. One very important goal of these detectors is to observe neutron stars and their mergers and to study their physical properties and equations of state. These detectors are expected to revolutionize the field of gravitational wave astronomy by achieving greater sensitivities over a larger bandwidth, thereby probing regions of the Universe at even larger distances than those currently achieved. In particular, signals with frequencies close to the inverse of the time spent by the light to make a round-trip within the arm of the currently planned interferometers (the so called Full-Spectral Range (FSR)), will experience an amplification that will make them observable in the frequency region dominated by shot-noise. The Cosmic Explorer, for instance, with an arm length of $40$ km, will display an FSR frequency of $3.75$ kHz at which its signal-to-noise ratio will be marked by a sharp maximum.\footnote{As will be shown in the following section, the amplification factor achievable by the response of a Fabry-Perot (FP) interferometer to a GW signal at the FSR frequency depends on the reflectivities of the test-mass mirrors.} The {\it w-modes}, first predicted by Kokkotas and Schutz \cite{KostasBernard92} and expected to be emitted by pulsating neutron stars in the kilohertz band, might therefore become observable by these detectors in this frequency region. The {\it w-modes} arise from the coupling of the oscillation of the star's matter to the gravitational wave oscillations of the space-time metric. Their detections will allow us to infer with high precision the mass and size of their sources \cite{Andersson_Kokkotas1998}, resulting in  very tight constraints on the equation of state.
\par
The article is organized as follows. In Section \ref{SecII} we provide the response of a FP interferometer to a GW signal of arbitrary wavelength. Taking advantage of the publicly available software programs that generate the anticipated displacement noise spectral densities of the Cosmic Explorer and Einstein Telescope projects, we then obtain the sensitivity curves for these two interferometers, which depend on the direction and the polarization of the signal. This result allows us to then derive in Section \ref{SecIII} an approximate analytic expression of the SNR of a damped-sinusoid, which describes these modes. We find that with the interferometer designs currently envisioned and for sources out to $0.8 {\rm Mpc}$ and emitting an energy that is $10^{-6}$ times the rest energy of the Sun, CE would achieve an average SNR of more than $5$, while ET will achieve an SNR of about $4$. In our computations, we have considered a  distance of $0.8$ Mpc which would reach out to the Andromeda galaxy and which consists of a trillion stars. This would essentially double the likely sources with those in the Milky Way. However, increasing the mirror reflectivities by a few percent over those currently envisioned would further reduce the noise spectrum at the FRS to a level that would result in an SNR $= 10$ for the two interferometers. 

\section{Response of a Fabry-P\'erot Interferometer to a Gravitational Wave Signal}
\label{SecII}
In a Michelson interferometer, whose arms of equal-length $L$ have at their ends proof-masses with reflecting mirrors acting as free falling test particles, laser light is introduced inside the two arms via a 50-50 beam-splitter. If we assume for the moment that light is reflected only once by the end mirrors before being recombined at a photo detector (while the process is run continuously), the effect of a weak gravitational wave train on the frequency of the coherent light appears at the photo detector four-times, namely at the time it is incident on the light source, at the two delayed distinct times when the wave interacts with the end proof-masses, and at the round-trip light time (delayed by $2L/c$) \cite{EW75,E84}. 

\begin{figure}[htbp]
\includegraphics[width = 7.0in, clip]{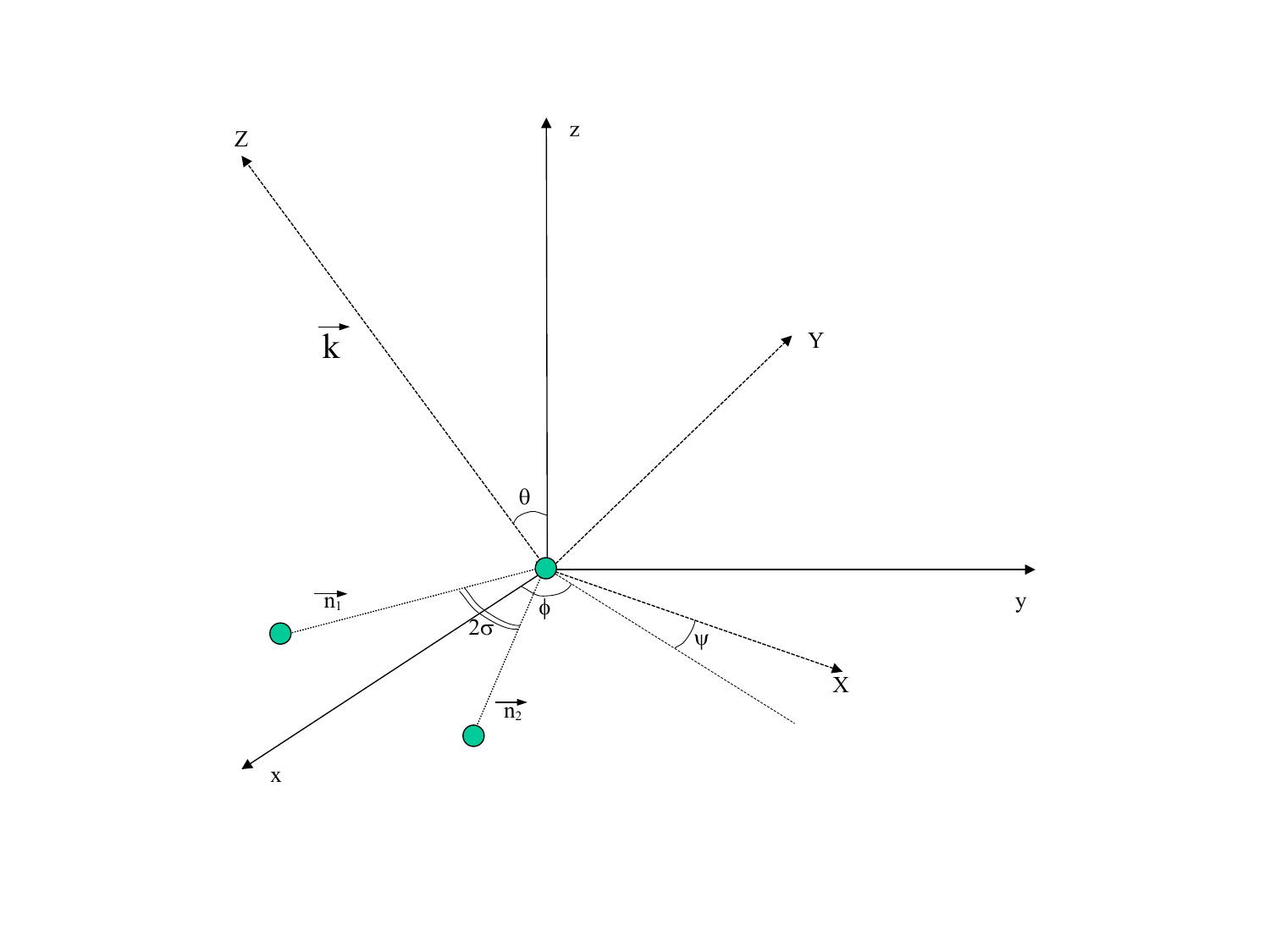}
\caption{The two coordinate systems associated with the interferometer and the GW signal. They are related through the Euler angles ($\theta, \phi, \psi$). See text for a complete description of the geometry.}
\label{Fig1}
\end{figure}

To explicitly write the response of a one-bounce Michelson interferometer to a gravitational wave signal, let us introduce a set of Cartesian orthogonal coordinates ($x, y, z$) centered on the corner proof-mass (see Fig. \ref{Fig1}). Here, the $x$-axis bisects the angle $2 \sigma$ enclosed by the two arms, while the $y$-axis is orthogonal to it in the plane of the interferometer; the $z$-axis is orthogonal to the plane of the interferometer and forms with ($x,y$) a right-handed orthogonal triplet. The coordinates associated with the wave instead, denoted by ($X,Y,Z$), are such that the $Z$-axis is aligned with the direction of propagation of the wave (represented by the unit vector $\vec k$). In the plane of the wave the two additional axes ($X,Y$) are orthogonal to each other and to the $Z$-axis, and the two components of the GW signal ($h_+, h_\times$) are defined with respect to them. We can then introduce the Euler angles ($\theta, \phi, \psi$) to relate the two coordinate systems when writing the interferometer response.

For an elliptically polarized gravitational wave signal, its polarization amplitudes  ($h_+, h_\times$) can be written in the form \cite{AET99}
\begin{equation}
    h_+(t) = H(t) \sin(\Gamma) \ \sin(2\pi f_0 t + \gamma) \ \ \ , \ \ \ 
    h_\times (t) = H(t) \cos(\Gamma) \ \sin(2\pi f_0 t) \ ,
\end{equation}
where $H(t)$ characterizes the strength of the gravitational wave and its frequency content does not include $f_0$. In addition, ($\Gamma, \gamma$) defines the wave  polarization state and are related to the coordinates in the Poincare sphere for spin-2 waves.

In the interferometer coordinate system the two-way Doppler responses from each arm can be written as follows
\cite{TA99} (units in which the speed of light $c = 1$).
\begin{eqnarray}
\left( \frac{\Delta \nu (t)}{\nu_0}\right)_1 \equiv 
y_1 (t) & = & 
\left[ - \frac{(1 - \vec k \cdot \vec {n}_1)}{2} \ \Psi_1 (t) \ - \ 
\vec k \cdot \vec {n}_1 \ \Psi_1 (t - (1 + \vec k \cdot \vec {n}_1)L) 
\right. \nonumber \\
& + &
\left. \frac{(1 + \vec k \cdot \vec {n}_1)}{2} \ \Psi_1 (t - 2L) 
\right]  \ ,
\label{eq:1}
\end{eqnarray} 
\begin{eqnarray}
\left( \frac{\Delta \nu (t)}{\nu_0}\right)_2 \equiv 
y_2 (t) & = & 
\left[ - \frac{(1 - \vec k \cdot \vec {n}_2)}{2} \ \Psi_2 (t) \ - \ 
\vec k \cdot \vec {n}_2 \ \Psi_2 (t - (1 + \vec k \cdot \vec {n}_2)L) 
\right. \nonumber \\
& + & \left. \frac{(1 + \vec k \cdot \vec {n}_2)}{2} \ \Psi_2 (t - 2L) 
\right]  \ ,
\label{eq:2}
\end{eqnarray} 
where $\vec k$ is the unit vector in the direction of propagation of
the planar gravitational wave pulse.  In Equations (\ref{eq:1},
\ref{eq:2}) we have denoted by $\vec {n}_1$, $\vec {n}_2$, the unit
vectors along arm $1$ and $2$, respectively, while we have denoted with $\Psi_{(1,2)} (t)$ the following two scalar functions
\begin{equation}
\Psi_{(1,2)} (t) = 
\left[ 
\frac{n^i_{(1,2)}n^j_{(1,2)}}
{
{1 - ({{\vec k} \cdot {\vec {n}}_{(1,2)}})^2}} \right] \ 
H_{ij} (t) \ ,
\label{eq:3}
\end{equation}
with the sum over repeated space-like indices assumed. In Eqs. (\ref{eq:3}) the components of the tensor $H_{ij} (t)$, the rank-2 tensor associated with the gravitational wave signal in the coordinate system $(x, y, z)$
\cite{TA99}, are functions of the three Euler angles ($\theta, \phi, \psi$), the two wave polarization amplitudes ($h_+(t), h_\times (t)$), and the polarization angles ($\Gamma, \gamma$) \cite{SchutzTinto}. From the above expressions of the effect of a GW signal on the frequency of light in each arm, the response of a one-bounce Michelson interferometer, $M (t)$, is equal to
\begin{equation}
    M(t) \equiv y_1(t) - y_2(t) \ .
\end{equation}
The reason for providing the above expression for a one-bounce Michelson interferometer is because it has been shown that the response functions of (i) a many-bounce Michelson Interferometer \cite{TE95} and (ii) a FP Michelson interferometer \cite{Rakhmanov_Romano_Whelan2008,Rakhmanov_2009} can be written in the Fourier domain as the product of the one-bounce response $M$, by a transfer function that accounts for the many round-trips made by the light within each arm before interfering at the photo detector. If we denote $M_{FP}(t)$ the relative frequency fluctuations of a FP interferometer that interacts with a GW signal, the expression of its Fourier transform is equal to  \cite{Rakhmanov_Romano_Whelan2008,Rakhmanov_2009}
\begin{equation}
    \widetilde{M}_{_{FP}}(f) \equiv \frac{\widetilde{M}(f)}{[1 - rr' e^{-4\pi i f L}]}
    \label{yfp} \ ,
\end{equation}
where $r$ and $r'$ are the reflectivities of the corner and end mirrors, respectively, and $f$ is the Fourier frequency. At the FSR frequency $f = 1/2L$, the detector response to a GW signal is amplified by the factor $(1-rr')^{-1}$. In the case of the Cosmic Explorer, ${\rm FSR} = 3.75$ kHz, while for a $20$ km Einstein Telescope configuration this is equal to $7.5$ kHz. The amplification of the GW signal at the FSR frequency, which is determined by the reflectivities of the mirrors, can be substantial in a frequency region where important scientific discoveries could be made.

\begin{figure}[htbp]
\includegraphics[width = 4.0in, clip]{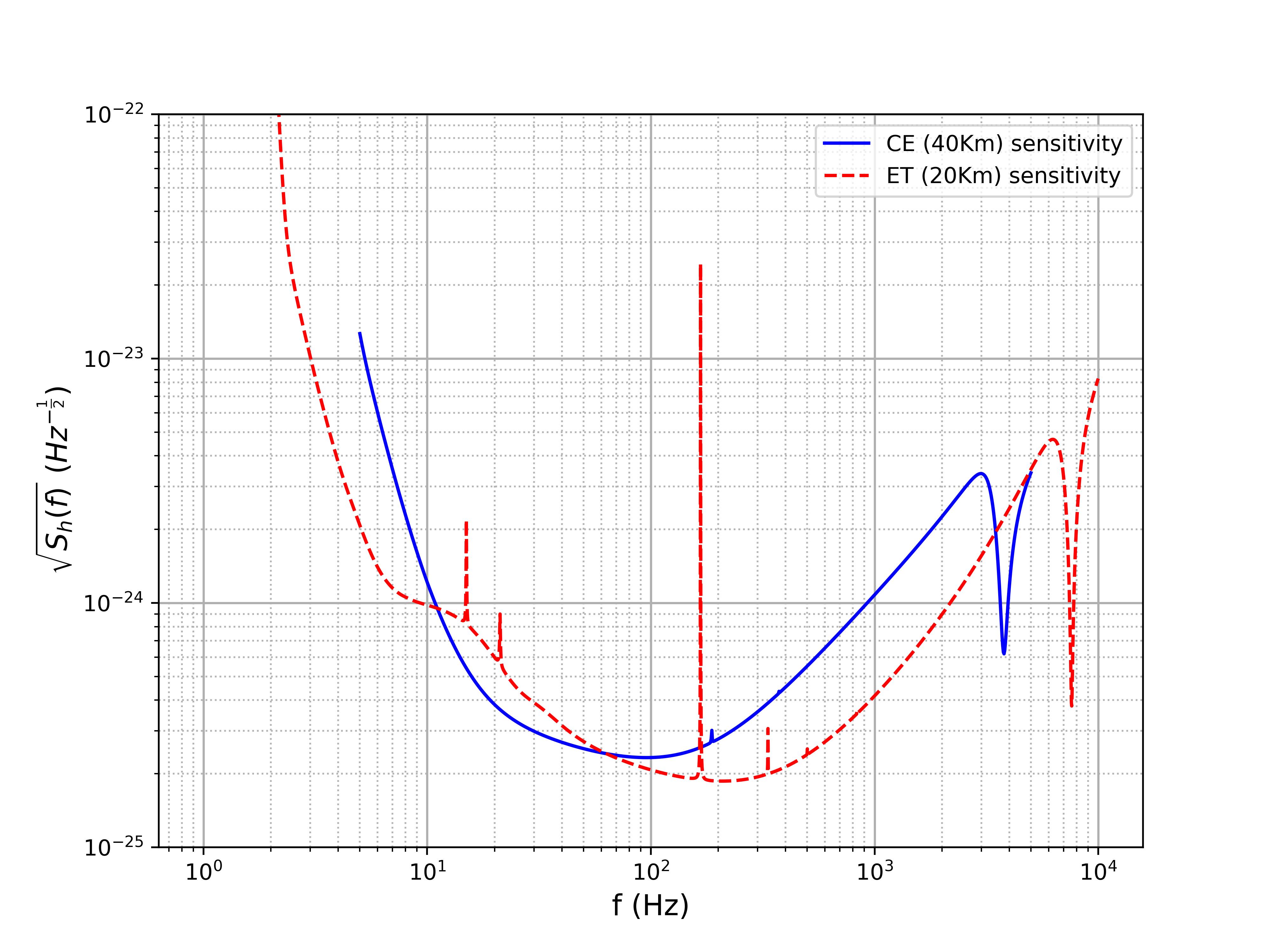}
\caption{Strain sensitivities of the Cosmic Explorer (CE) and Einstein Telescope (ET) projects. The CE project envisions an arm length of $40$ km, while current sensitivity studies for ET have been considering arm lengths ranging from $10$ to $20$ km. Here we have assumed an ET arm length of $20$ km, which results in a FSR frequency of $7.5$ kHz.}
\label{fig_CE_asd}
\end{figure}

Following technical studies on expected noises associated with the designs of the two projects \cite{CE_1,CE_2,ET_1}, in figure \ref{fig_CE_asd} we plot the strain sensitivities of both CE and ET, where for ET we have assumed an arm length of $20$ km. Since the information provided about the expected performance of the two projects was in different formats, their corresponding sensitivity curves were derived differently. In the case of the CE project, the expected interferometer displacement spectral density was made available in the form of an ASCII file, which already included the FP GW signal transfer function  $[1 - rr' e^{-4\pi i f L}]$. Therefore, to obtain the FP interferometer strain sensitivity to an ensemble of GW signals of random location in the sky and random polarization states, we first had to convert the provided displacement spectral density to the corresponding Doppler spectral density. Then we calculate numerically the r.m.s. of the one-bounce Michelson response transfer function taken over GW signals randomly distributed over the celestial sphere and with random polarization states. Its expression is equal to
\begin{equation}
    T(f) \equiv \sqrt{\Big\langle \frac{ |{\widetilde M}(f)|^2}{ {|\widetilde A}(f)|^2} \Big\rangle_{\theta,\phi,\psi,\Gamma,\gamma}}
    \label{T}
\end{equation}
Finally, we divide the square-root of the Doppler spectral density of the noise by the function $T(f)$ given in Eq. (\ref{T}) to obtain the CE strain sensitivity curve (solid line) plotted in Fig.(\ref{fig_CE_asd}).

In the case of ET sensitivity, we use the online program {\it Plot Digitizer} \cite{WebPlotDigitizer} to extract the strain sensitivity provided in Figure 1 of reference \cite{ET_1} valid for a configuration with arm length $20$ km (see the dashed-line in Fig.(\ref{fig_CE_asd}).

\section{Approximate analytical expression of the SNR of a GW mode}
\label{SecIII}

\subsection{The general expression for the SNR}

We model the GW mode as a damped sinusoid with damping constant $\lambda$ and frequency $\om0$. The damping time or the e-folding time $\tau$ is related to $\lambda$ by the relation $\tau = 1/\lambda$. We also place the beginning of the signal at $t = 0$. Thus, we write the GW mode $h(t)$ as: 
\begin{equation}
h (t) = \begin{cases}   
    A e^{- \lambda t} \cos {\om0 t }~~~~~~~~~~~~~ t \geq 0 \,, \\
  0  ~~~~~~~~~~~~~~~~~~~~~~~~~~~~~t < 0 \, ,
  \end{cases}
\end{equation}
where $A$ is the constant amplitude of the signal. Its Fourier transform is given by
\be
\th (f) ~=~ \int_{-\infty}^{\infty} ~h(t) ~e^{- 2 \pi i ft} ~dt \,.
\label{eq:Fourier-f}
\ee
or,
\be
\th (\omg) ~=~ \int_{-\infty}^{\infty} ~h(t) ~e^{- i \omg t} ~dt \,.
\label{eq:Fourier_omega}
\ee
The usual practice is to use the one-sided spectral density (PSD) $S_h (f)$, where the negative frequency components are folded over the positive frequency components. Then if we match filter the signal, the square of the SNR $\rho$ is given by the equation
\be
\rho^2 = 4~\int_{0}^{\infty} ~\frac{|\th (f)|^2}{S_h (f)}~df = \frac{2}{\pi} \int_{0}^{\infty} ~\frac{|\th (\omg)|^2}{S_h (\omg)}~d \omg\,.
\ee


In a typical GW mode, $\lambda \ll \om0$; for example, $\tau \sim 0.1$ sec, while $f_0$ is few kHz and thus $\om0 \sim 10^4$ rad/sec. In this limit, the band-width of the mode is very small $\sim \lambda$ and so it is typically a narrow band signal. Over this small band-width the PSD $S_h (f)$ remains essentially constant and can be pulled out of the integral. We will justify this in subsection \ref{subsec:bandwidth}. Thus, we may write
\be
\rho^2 = \frac{4}{S_h (f_0)} \int_{0}^{\infty} ~|\th (f)|^2 ~ df \ .
\ee
Our task is to compute the integral in order to obtain the SNR, and this can easily be done in the time domain. In order to achieve this, we rely on Parseval's theorem. We write
\be
I =  \int_{- \infty}^{\infty} ~|\th (f)|^2 ~ df = \int_{-\infty}^{\infty} ~|h (t)|^2 ~ dt \,.
\ee
The integral in the time domain can easily be performed - the integration range is $t \geq 0$, because the signal is zero when $t < 0$. Thus,
\bea
I &=& A^2 \int_0^{\infty}~ e^{-2 \lambda t}~\cos^2 \om0 t ~ dt \, \no \\
  &=& A^2 \times \qtr \left [ \frac{1}{\lambda} + \frac{\lambda}{\lambda^2 + \om0^2} \right ] \,.
 \label{eq:time_domain_power} 
\eea
When $\lambda << \om0$, the second term can be neglected, and therefore we obtain $I \simeq A^2/4 \lambda$. Then the SNR $\rho$ is given by
\be
\rho \approx \frac{A}{\sqrt{2 \lambda}} \left[ \sqrt{S_h (f_0)} \right]^{-1} = A \sqrt{\frac{\tau}{2}} \left[ \sqrt{S_h (f_0)} \right]^{-1} \,.
\label{eq:SNR}
\ee
This result is clear if we observe that the integral of $e^{-2 \lambda t}$ in the range $[0, \infty]$ is $1/2 \lambda$ and that the term $\cos^2 \om0 t$ averages to $1/2$ and so the integral is approximately $1/4 \lambda$.   
\vss

\subsection{The band-width of the GW mode}
\label{subsec:bandwidth}

We also compute the Fourier transform of the GW mode in order to explicitly  ascertain its band-width. We  are especially interested in the case where $\lambda << \om0$, where we expect the bandwidth to be small enough so that the approximation we have made in computing the SNR is valid. Since the band-width does not depend on the amplitude $A$, we set $A = 1$.
In the Fourier domain, we have, from Eq. (\ref{eq:Fourier_omega}),
\be
\th (\omg) = \frac{\lambda + i \omg}{(\lambda + i \omg )^2 + \om0^2} \,,
\ee
and therefore,
\be
|\th (\omg) |^2 = \frac{\lambda^2 + \omg^2}{[\lambda^2 - (\omg^2 - \om0^2)]^2 + 4 \lambda^2 \omg^2} \,.
\ee
Note that $|\th (\omg)|^2$ is an even function of $\omg$, that is, $|\th (\omg) |^2 = |\th (-\omg) |^2$. We therefore evaluate the integral for positive frequencies only and then multiply the result by a factor of 2. In the limit $\lambda << \om0$, we set $\omg = \om0 + \Dom$, where $\Dom << \om0$. Then in this limit,
\be
|\th (\omg) |^2 \approx \frac{1}{4(\lambda^2 + \Dom^2)} \,.
\label{eq:Lorentzian}
\ee

\begin{figure}[t]
    \centering
    \includegraphics[width=0.48\textwidth,clip]{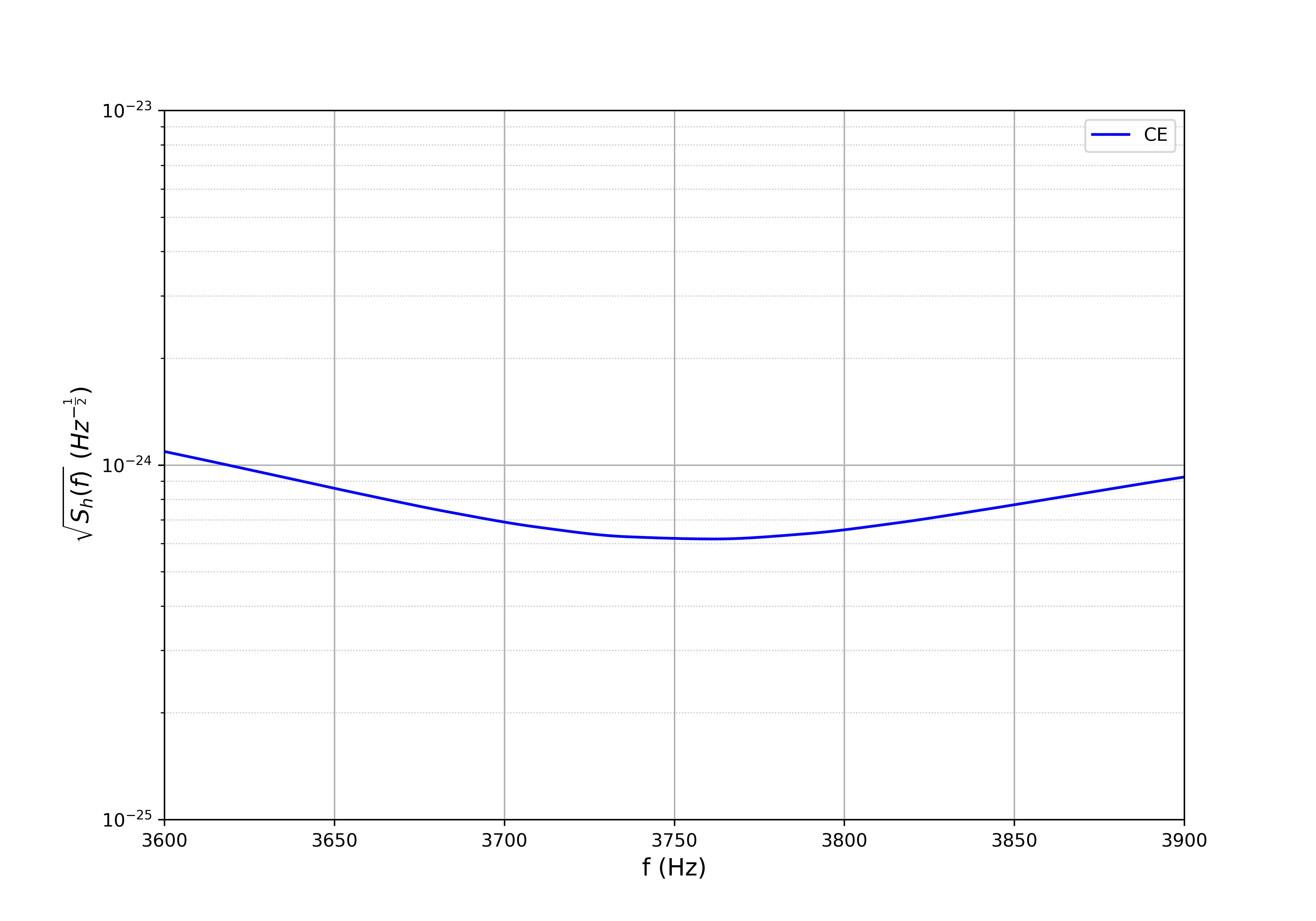}
    \hfill
    \includegraphics[width=0.48\textwidth,clip]{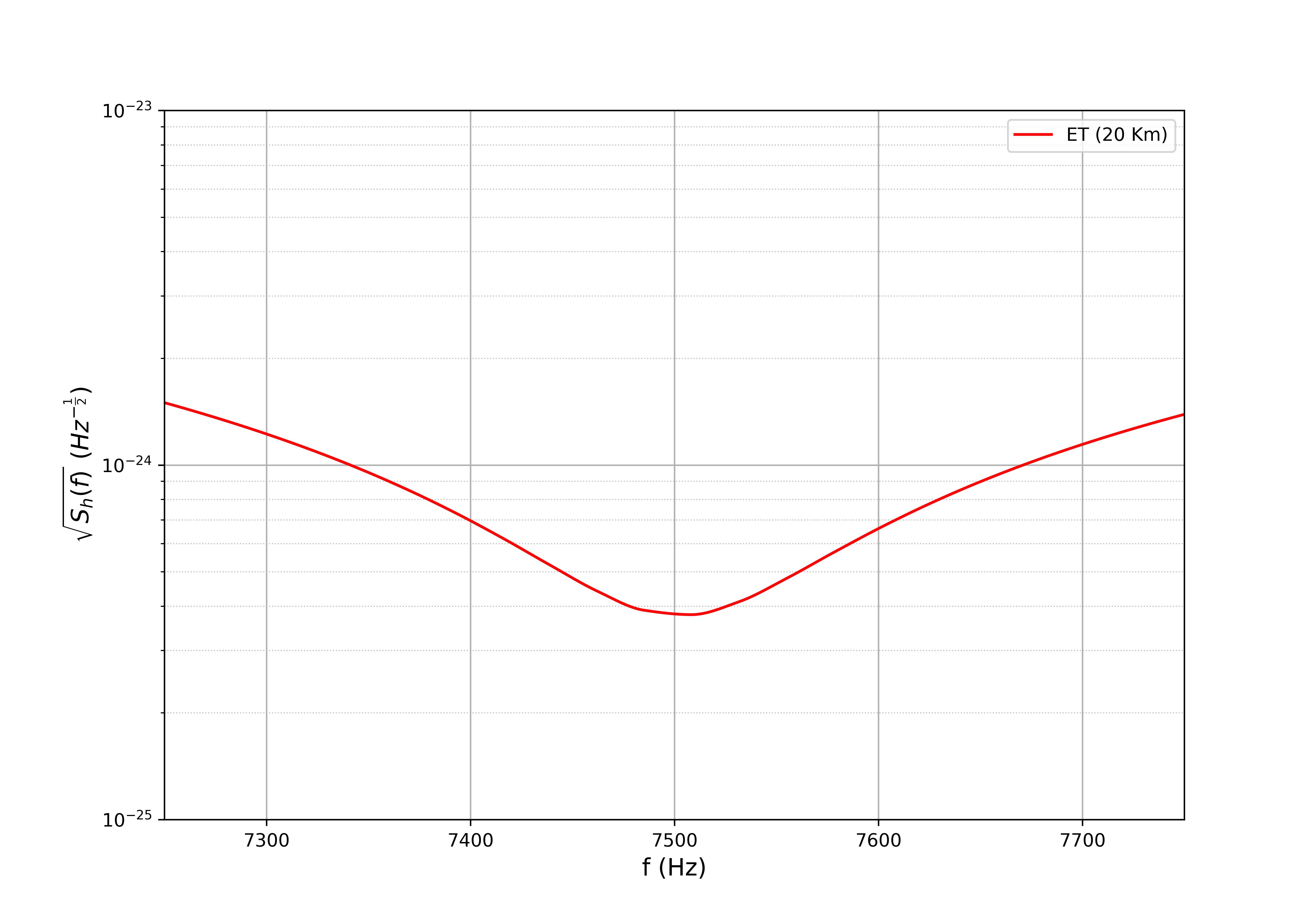}
    \caption{Magnified view of the CE (40 Km) and ET (20 Km) sensitivity curve highlighting the flatness of the response in the free spectral range (FSR) region.}
    \label{fig_ce_fsr_asd_mag}
\end{figure}

\begin{figure}[htbp]  \centering
    \begin{subfigure}{0.45\textwidth}
       \centering
        \includegraphics[width=\linewidth]{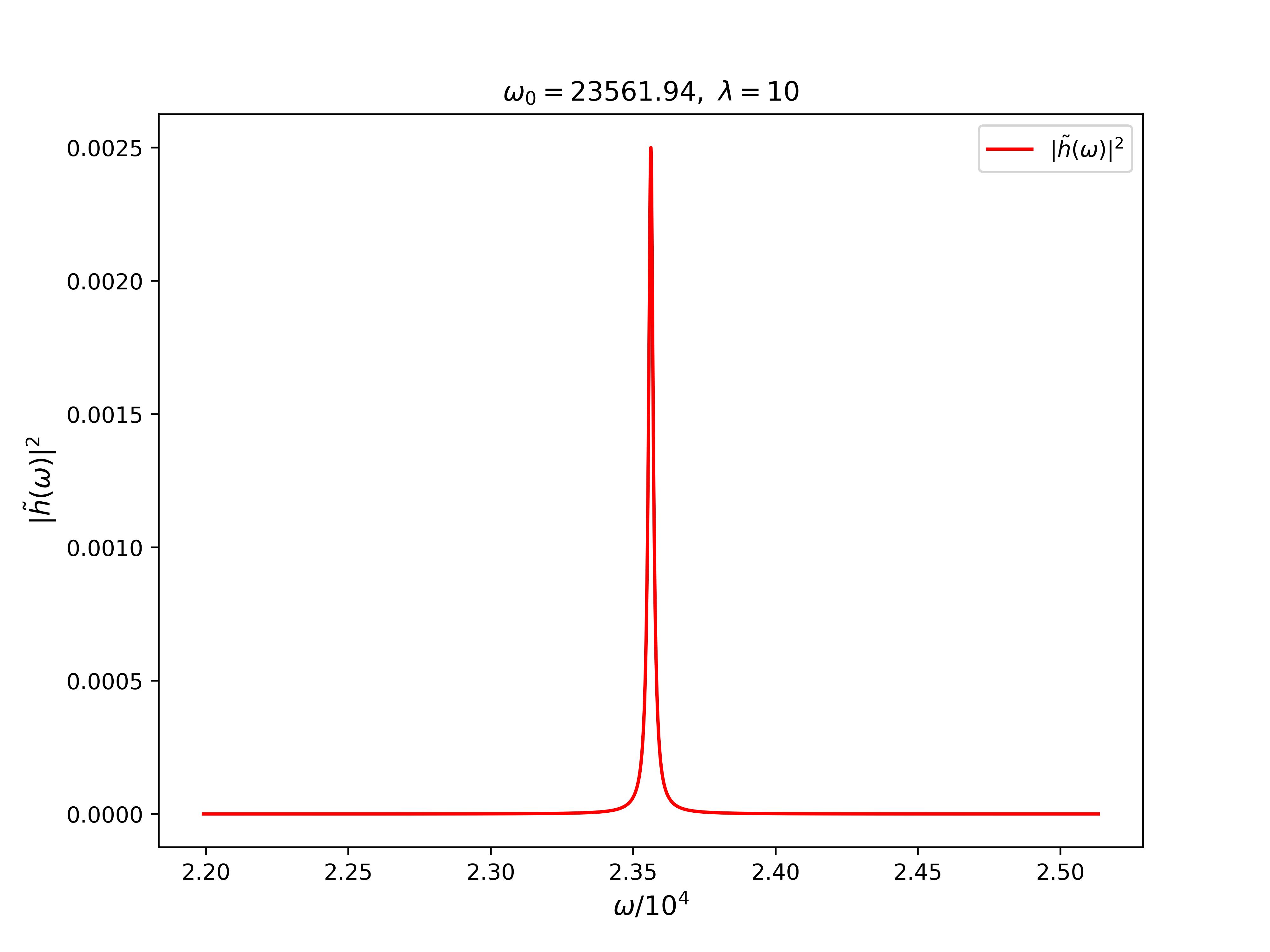}
        \phantomcaption
        \label{fig_h_sq}
        \raisebox{1.5ex}{\makebox[0pt][l]{\textbf{(a)}\hspace{0.5em}}}
    \end{subfigure}
    \hfill
    \begin{subfigure}{0.45\textwidth}
        \centering
        \includegraphics[width=\linewidth]{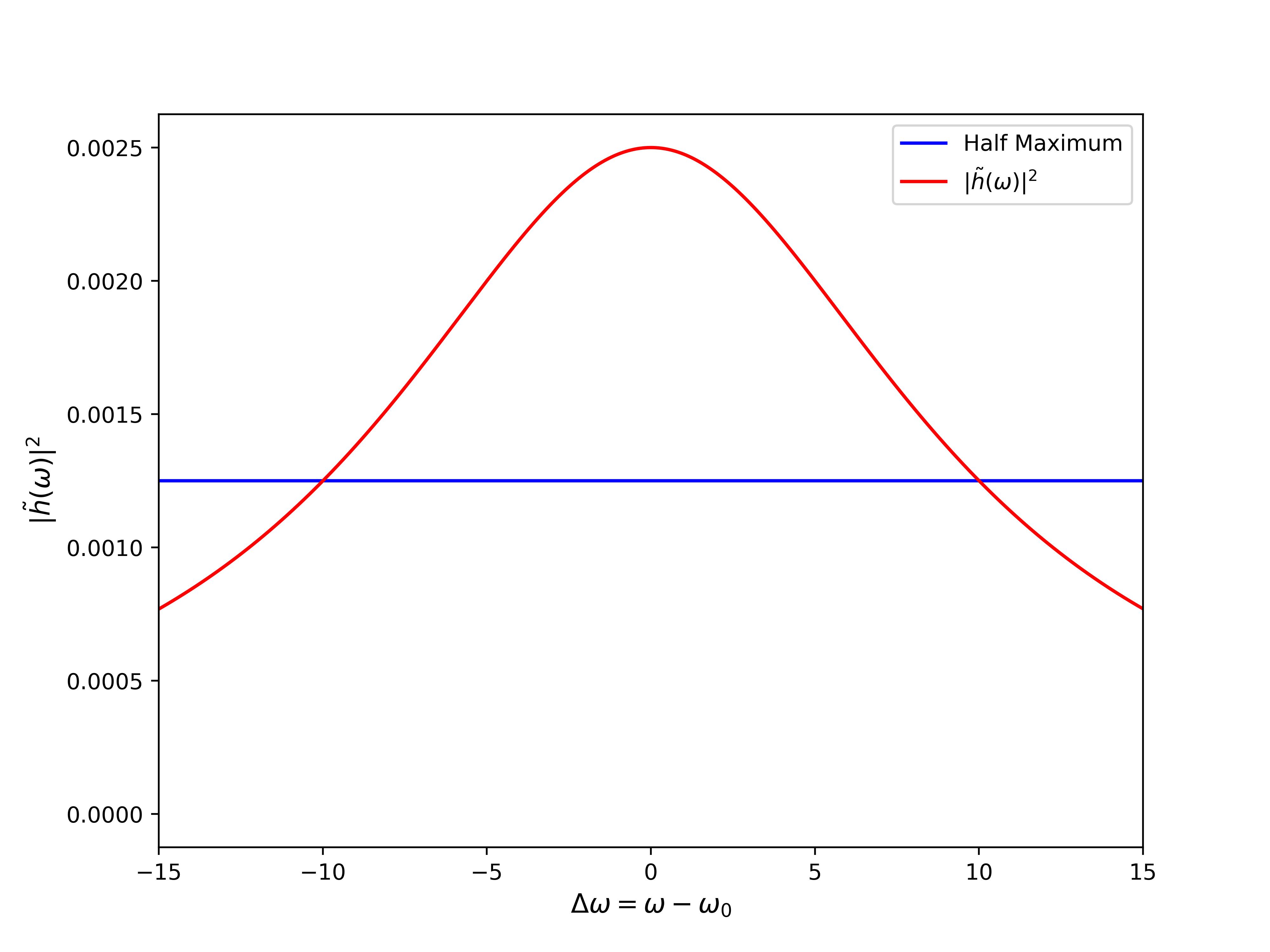}
        \phantomcaption
        \label{fig_h_sq_fwhm}
        \raisebox{1.5ex}{\makebox[0pt][l]{\textbf{(b)}\hspace{0.5em}}}
    \end{subfigure}
    \caption{
    Fig. (a) shows $|\tilde{h}(\omega)|^2$ as a function of $\omega$ with $A = 1$ and $\lambda = 10$. As seen from Eq. (\ref{eq:time_domain_power}) the height of the peak is $\sim 1/4 \lambda = 0.025$. In Fig (b) a zoomed version of figure (a) is shown in the vicinity of the peak and it further displays the FWHM $\simeq 2 \lambda = 20$ rad/sec. which corresponds to about 3 Hz.}
    \label{fig_h_sq_w_fwhm}
\end{figure}

This is a Lorentzian and its integral from $-\infty$ to $\infty$ over $\Dom$ is $\pi/ 4 \lambda$. This quantity must be multiplied by 2 and divided by $2 \pi$, and therefore this result agrees with the integral in the time domain in Eq. (\ref{eq:time_domain_power}) when $A = 1$ and in the limit $\lambda << \om0$. The full width at half maximum (FWHM) of the Lorentzian occurs when $\Dom = \lambda$ and is therefore $2 \Dom = 2 \lambda$. This is evident from Fig. \ref{fig_h_sq_w_fwhm}. Most of the SNR is captured within a range of a few $\lambda$. For example, 90 \% of the SNR is obtained by integrating up to about  $6 \lambda$ which  translates to about $\Delta f \sim 10$ Hz for $\tau \sim 0.1$ sec. This means that the integration range is $\sim 20$ Hz. Over this range the PSD is essentially constant or flat - the change is within a few percent. This is seen from Fig. \ref{fig_ce_fsr_asd_mag}. Therefore, the approximation we have made is justified.

\newpage

\subsection{The amplitude and SNR of the GW mode}

In order to compute the SNR Eq. (\ref{eq:SNR})
we need the value of the amplitude $A$ of the mode. We generally have an estimate of the energy that is radiated out by the mode. For a supernova explosion the energy is $E \sim 10^{-6} \Msun$. We therefore relate the amplitude $A$ to the energy $E$. For this, we need to integrate the flux of GW of the mode over all directions. The flux of a monochromatic GW of frequency $\omg$ is related to the GW amplitude $h$ by the equation:
\be
\F = \frac{c^3}{16 \pi G} \omg^2 h^2 \,,
\ee
where $\F$ denotes the flux of the GW in a specific direction. Thus, the energy of a GW mode with amplitude $A$ is given by integrating the flux over time and over a sphere of radius $r$. This results in the following
\be
E = \frac{c^3}{16 \pi G} \omg^2 \int_{-\infty}^{\infty} dt~ |h(t)|^2  \times 4 \pi r^2 \,.
\label{eq:energy}
\ee
However, the flux is not uniform in all directions and its average over all directions is relevant. But here we need not consider this, because it has already been taken into account in the sensitivity curve as described in section \ref{SecII}.  
\vss
The integral appearing in Eq. (\ref{eq:energy}) is $\sim A^2 /4 \lambda = A^2 \tau/4$ when $\om0 \tau >> 1$. Putting all this together, we obtain the following
\be
E = \frac{c^3}{16 G} \om0^2 A^2 \tau r^2 \,.
\ee
The above equation may be inverted for $A$ and the result is:
\be
A = \frac{1}{2 \pi} \sqrt{\frac{16 G}{ c^3}} 
f_0^{-1} \tau^{-1/2} E^{1/2} r^{-1} \,.
\label{eq:amplitude}
\ee
where we have written $\om0 = 2 \pi f_0$. We now estimate the amplitude $A$ of the mode for the Cosmic Explorer. We take the distance $r$ to $0.8$ Mpc in order to include likely sources in the Andromeda galaxy which contains of the order of a trillion stars. Therefore,
\be
A_{\rm CE} = 1.45 \times 10^{-23} \left [\frac{f_0}{3.75 ~{\rm kHz}} \right]^{-1} \left[ \frac{\tau}{0.1~{\rm sec}} \right]^{-1/2} \left[ \frac{r}{0.8 ~{\rm Mpc}} \right]^{-1} \left[\frac{E}{10^{-6} \Msun c^2} \right]^{1/2} \,. 
\ee

For ET, since the FSR frequency is double that of CE, the corresponding amplitude is halved, keeping all other parameters the same.

Combining the expression for the SNR $\rho$ given in Eq. (\ref{eq:SNR}) with the amplitude $A$ given in Eq. (\ref{eq:amplitude}) we obtain the expression for the SNR as:
\be
\rho = \frac{1}{2 \pi} \sqrt{\frac{8 G}{c^3}} f_0^{-1} r^{-1} E^{1/2} [S_h (f_0)]^{-1/2} \,.
\ee
For the Cosmic Explorer, we have $f_0 = 3.75$ kHz and for the ASD, we consider the plot in Fig \ref{fig_CE_asd}. Therefore, the typical SNR, which we denote by $\rho_{\rm CE}$, is given by
\be
\rho_{\rm CE} = 5.41 \times \left [\frac{f_0}{3.75 ~{\rm kHz}} \right]^{-1}  \left[ \frac{r}{0.8~{\rm Mpc}} \right]^{-1} \left[\frac{E}{10^{-6} \Msun c^2} \right]^{1/2} \left[\frac{\sqrt{S_h (f_0)}}{6 \times 10^{-25}~{\rm Hz}^{-1/2}} \right]^{-1} \,. 
\label{eq:SNR_CE}
\ee

In case of the Einstein Telescope, we have $f_0 = 7.5$ kHz and the ASD is $\sim 4 \times 10^{-25}$ Hz$^{-1/2}$ in the relevant frequency range, so we obtain,
\be
\rho_{\rm ET} = 4.05 \times \left [\frac{f_0}{7.5 ~{\rm kHz}} \right]^{-1}  \left[ \frac{r}{0.8~{\rm Mpc}} \right]^{-1} \left[\frac{E}{10^{-6} \Msun c^2} \right]^{1/2} \left[\frac{\sqrt{S_h (f_0)}}{4 \times 10^{-25}~{\rm Hz}^{-1/2}} \right]^{-1} \,.
\label{eq:SNR_ET}
\ee
Although the statistical significance of the above SNRs is on the borderline of detection, several considerations should be made about the above SNR estimates. First, they are average values taken over source sky location and polarization states. As such, there will be signals that emit the selected amount of energy that will result in SNRs higher than their averages. Second, we assumed a somewhat conservative amount of emitted energy. Taking it to be ten times higher, the SNR would increase by a factor of $\sqrt{10}$, resulting in a statistically significant SNR for the two detectors. Third, as a possible instrumental modification, one could select mirrors with higher reflectivities to further reduce the value of the noise spectra at the FSR frequency. We have estimated that increasing the product $rr'$ of reflectivities to $92 \%$ would result in SNR $= 10$ in CE, while bringing $rr'$ to $97\%$ would also result in a SNR $= 10$ for ET.

\section{Concluding Remarks and Summary}
\label{SecIV}

 Most searches for GW sources focus on the wide portion of the bandwidth of  the sensitivity curves in the context of the third generation detectors such as the Cosmic Explorer and Einstein Telescope. The frequency range considered typically spans from a few Hz to kHz or little more. On the other hand, in this work we have focused on the high frequency range of these detectors, namely, the band of frequencies in the neighbourhood of the FSR frequencies. Because of their long arm-lengths, their FSR frequencies are relatively lower, namely, 3.75 kHz for the Cosmic Explorer and 7.5 kHz for the 20 km Einstein Telescope and fall in the range of frequencies emitted by the GW modes. We have investigated the detectability of high frequency sources, specifically the GW modes emitted by pulsating neutron stars and the {\it w-modes} due to the coupling between the oscillations of the star's matter to the gravitational wave oscillations of the space-time metric. 
 \par
 We find that the modes have sufficiently narrow band-width over which the noise PSD does not change appreciably and this allows for convenient evaluation of the SNR in the time domain using Parseval's identity. Furthermore, our investigations show that the computed SNRs are significant if the sources are located within 0.8 megaparsec, which includes the Andromeda galaxy, and further, the associated energy is ten times larger than that assumed in Eq.(\ref{eq:SNR_CE}) and Eq.(\ref{eq:SNR_ET}), then the respective SNRs are enhanced by a factor of $\sqrt{10} \simeq 3.16$ - the SNR for the Cosmic Explorer would be enhanced to about 17 and that for the Einstein Telescope to about 13. In addition, we have shown that a minor increase in the reflectivities of the currently envisioned mirrors would result in a SNR of $10$ for both interferometers at their FSR frequencies.

\section*{Acknowledgments}

The authors thank Dr. Kevin Kuns for providing technical data on the Cosmic Explorer project, and acknowledge Professor Yanbei Chen for pointing out the work done on the subject. For M.T., this research was funded by the Polish National Science Center Grant No.  2023/49/B/ST9/02777. M.T. thanks the National Institute for Space Research (INPE, Brazil) for their kind hospitality while this work was done. 
\bibliographystyle{apsrev}
\bibliography{refs}
\end{document}